\documentclass[reprint,superscriptaddress,amsmath,amsfonts,amssymb,aps,prl,showkeys]{revtex4-1}
\usepackage{bm}
\usepackage{color}
\usepackage{graphicx}
\usepackage[normalem]{ulem}
\usepackage[makeroom]{cancel}

\newcommand{\Rs}{$R_\text{sw}$ }
\newcommand{\dg}{^\circ }

\usepackage{hyperref}
\hypersetup{
  breaklinks = {true},
  citecolor = {blue},
  colorlinks = {true},
  linkcolor = {red},
  urlcolor = {blue},
}

\begin{document}

\title{Giant Resistance Switch in Twisted Transition Metal Dichalcogenide Tunnel Junctions}

\author{Marc Vila}
\affiliation{Department of Physics, University of California, Berkeley, California 94720, USA}
\affiliation{Materials Sciences Division, Lawrence Berkeley National Laboratory, Berkeley, California 94720, USA}


\date{\today}

\begin{abstract}

Resistance switching in multilayer structures are typically based on materials possessing ferroic orders. Here we predict an extremely large resistance switching based on the relative spin-orbit splitting in twisted transition metal dichalcogenide (TMD) monolayers tunnel junctions. Because of the valence band spin splitting which depends on the valley index in the Brillouin zone, the perpendicular electronic transport through the junction depends on the relative reciprocal space overlap of the spin-dependent Fermi surfaces of both layers, which can be tuned by twisting one layer. Our quantum transport calculations reveal a switching resistance of up to $10^6 \%$ when the relative alignment of TMDs goes from $0\dg$ to $60\dg$ and when the angle is kept fixed at $60\dg$ and the Fermi level is varied. 
By creating vacancies, we evaluate how inter-valley scattering affects the efficiency and find that the resistance switching remains large ($10^4 \%$) for typical values of vacancy concentration. Not only this resistance switching should be observed at room temperature due to the large spin splitting, but our results show how twist angle engineering and control of van der Waals heterostructures could be used for next-generation memory and electronic applications.

\end{abstract}

\maketitle


\textit{Introduction ---} A two-level resistance switch is at the core of information storage and processing in current technologies, and continues to be nowadays a major field of research \cite{Wang2020, Lanza2022}. Typically, this phenomenon is achieved in multilayer structures comprised of a metal-insulator-metal junction, with different mechanisms responsible for the resistance variation such as ferroelectric polarization \cite{Zhuravlev2005, Garcia2014}, spin-dependent tunneling \cite{Moodera1995, Parkin2004, Yuasa2004} or ion migration \cite{Waser2007, Lanza2019}. The high and low resistance states can be associated with a bit of information, and therefore, the larger the difference between them, the more reliably those states can be distinguished between each other. The resistance switching can be quantified as the change in percentage between the high ($R_\text{high}$) and low ($R_\text{low}$) resistance values,
\begin{equation}
    R_\text{sw} = \frac{R_\text{high} - R_\text{low}}{R_\text{low}} \times 100\%.
\end{equation}
In magnetic tunnel junctions, for example, $R_\text{low}$ and $R_\text{high}$ correspond to metals with parallel and antiparallel magentic orientations, and \Rs typically reaches values of few hundreds \cite{Dieny2017}. 

Although different physics govern the distinct types of switching devices, common challenges exist that limit their potential. Interface quality and scalabilty are major concerns, especially in ferroic systems where decreasing the system size may render the ordered phases more susceptible to thermal fluctuations \cite{Dieny2017, Zhang2021, Cheng2022} and interface details can impact the transport and polarization properties \cite{Tsymbal2007, Burton2011, Garcia2014}. For that reason, two-dimensional materials have attracted a lot of attention as a way to circumvent those problems given their intrinsic low dimensionality and the ability to form pristine interfaces when stacked in van der Waals heterostructures  \cite{Geim2013, Hui2017, Zhang2021, Yang2022, Xue2022}, and they have already shown exceptional switching capabilities of up to $10^6 \%$ \cite{Song2018, Klein2018, Wang2018, Wang2018NC, Kim2018, Li2019, Wu2020, Wang2022, Xie2023, Zhu2023}. 

Another recent improvement in the field of magnetic tunnel junctions has been the fact that momentum-dependent transmission between the different layers involved in the vertical transport \cite{Karpan2007, Karpan2008} could realize resistance switches in many antiferromagnets, therefore debunking the idea that only ferromagnets possess spin-filtering properties. The variety of antiferromagnetic orders range from collinear \cite{Shao2023} and noncolinear \cite{Dong2022, Qin2023, Chen2023} to nonrelativistic spin-split antiferromagnets \cite{Shao2021, Smejkal2022TMR} (so-called altermagnets \cite{Smejkal2022, Smejkal2022Review}), and the generality of this momentum-dependent transmission is such, that it was also suggested for in-plane transport in transition metal dichalcogenide (TMD) monolayers \cite{Pulkin2016} and twisted graphene multilayers between Cu or Ni \cite{Hallal2020}. Such effect, however, does not need to be restricted to magnetic materials and should appear generally in systems possessing momentum-dependent spin splittings, such as materials with strong spin-orbit coupling (SOC). 

In this Letter, we apply the concept of momentum-dependent transmission in tunnel junctions comprised of TMD monolayers to realize a resistance switch based on their valley-dependent spin splitting \cite{Xiao2012}. The switching is achieved by twisting one of the layers by $60\dg$, which effectively reverses its valley-dependent spin splitting resulting in a high resistance state. By using quantum transport calculations we obtain \Rs larger than $10^6 \%$ between twist angles $\theta = 0\dg$ and $\theta = 60\dg$. Such \Rs can also be attained at a constant relative angle of $60^\circ$ with just tuning the Fermi level. Not only the predicted effect should be present at room temperature given the large spin-orbit splitting, but the efficiency remains as large as $10^4 \%$ for typical values of vacancy concentration, suggesting a very clear experimental signature that could be probed with recent methods of twist angle control \cite{RibeiroPalau2018, Inbar2023}.

\begin{figure}[tb]
\includegraphics[width=0.4\textwidth]{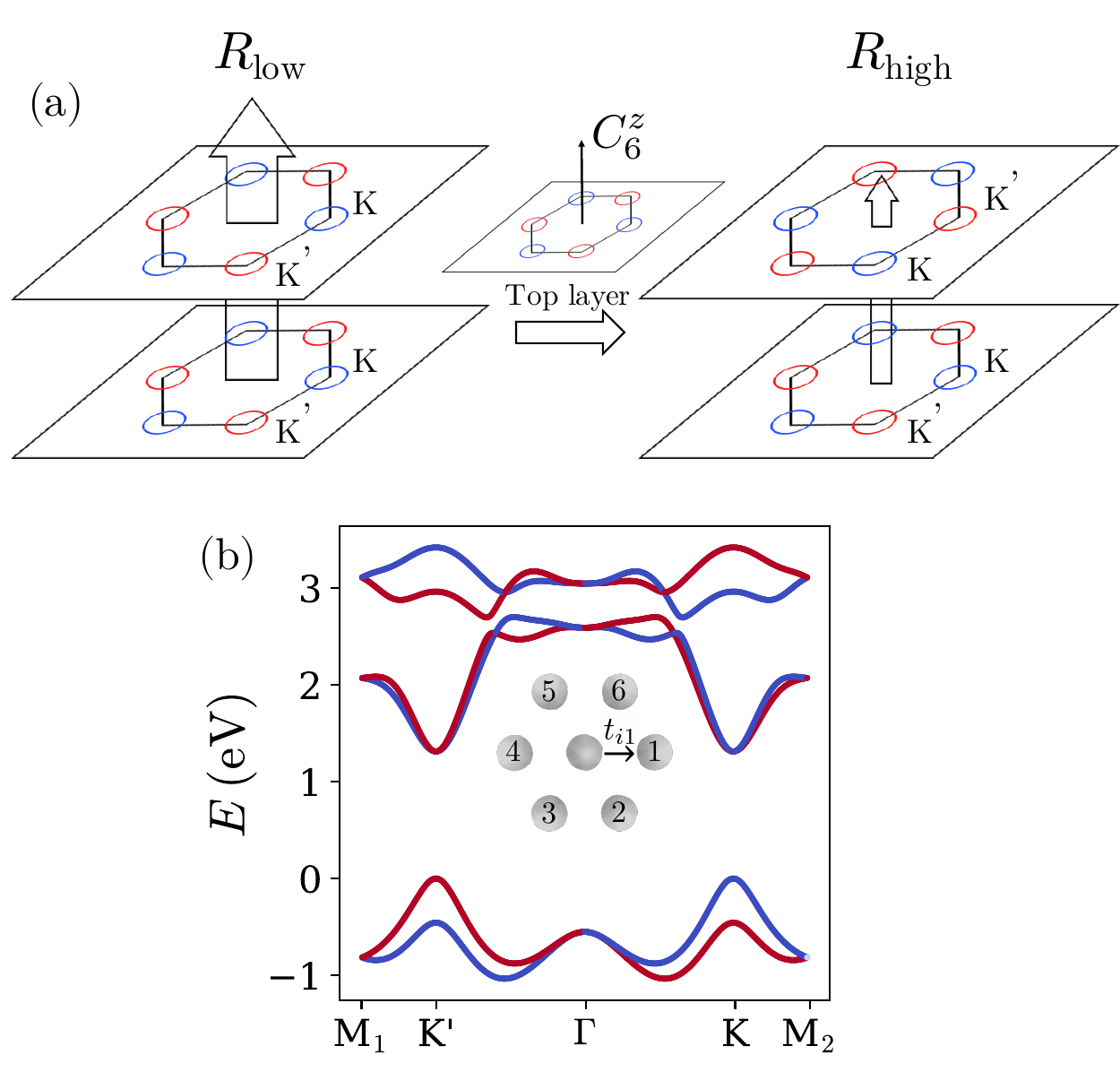}
\caption{(a) Schematics of the switching resistance based on twisted TMD tunnel junction, where the tunnel barrier has been omitted for clarity. When the spin-valley locking is the same in both layers, transport accros the junction leads to a low resistance state, $R_\text{low}$. In contrast, rotating one of the layers by $60\dg$ (i.e. $C_6^z$ operation) leads to an effective swapping of the spin-valley locking sign in that layer, which impedes vertical transport and results in a large resistance, $R_\text{high}$. (b) Band structure of the tight-binding model Eq. \eqref{eq_TMD} with the parameters of WSe$_2$ monolayer \cite{Liu2013}. The zero energy is set at the valence band maximum. Inset: triangular lattice of the model with the one of the nearest-neighbor hopping.}
\label{fig_F1}
\end{figure}

One of the most widely known property of hexagonal TMDs is the spin-valley locking of the valence bands that produces an opposite spin splitting at $K$ and $K^\prime$ valleys \cite{Zhu2011, Xiao2012}. This is described by the effective SOC term $\Delta s \tau$, with $2\Delta$ being the splitting, $s$ the spin index ($+1$ and $-1$ for up and down spins, respectively) and $\tau$ the valley index ($+1$ and $-1$ for $K$ and $K^\prime$, respectively). The splitting reaches up to few hundreds of meV, which means that each valley is fully spin polarized at low energies. In principle, the sign of the spin-valley locking $s \tau$ is fixed for a given material, and determines whether e.g. $K$ valley is up or down spin-polarized. Nevertheless, since a symmetry operation applies to both real and reciprocal space, one can take a TMD bilayer and rotate only one of the layers with the result being two Brillouin zones twisted with respect to one another. Due to the three-fold rotation symmetry of TMDs, if the twist angle $\theta$ is $0\dg$ plus a multiple of $120\dg$, the bilayer stack remains the same. However, for twist angles of $60\dg$ plus a three-fold rotation, $K$ and $K^\prime$ are interchanged and consequently the sign of $s \tau$ is effectively reversed, resulting in opposite spin splittings (between top and bottom TMDs) sharing the same momentum space in the common Brillouin zone. This procedure is illustrated in Fig. \ref{fig_F1}(a), and was used to control the excitonic properties in bilayer WSe$_2$ \cite{Jones2014} and to generate a quantum anomalous Hall insulator in MoTe$_2$/WSe$_2$ heterobilayers \cite{Li2021}.

If we now separate this bilayer with a thin insulating barrier like hexagonal boron nitride, the transport through the junction will just involve tunneling of TMD states between top and bottom layers. Therefore, the tunneling conductance or resistance will strongly depend on whether those states have the same or opposite spin polarization given by $s \tau$, akin to ferromagnetic tunnel junctions where that role is given by the magnetization. In this way, when $\theta = 0\dg$, both TMDs have effectively the same $s \tau$ and the transport through the junction will show a low resistance state. In contrast, we expect a large resistance increase for $\theta = 60\dg$ due to swapping $s \tau$ in one of the layers. This idea is shown in Fig. \ref{fig_F1}(a) and is the core concept of our work, which we now substantiate with quantum transport numerical calculations.

\textit{Device modeling ---} To calculate transport properties in a tunnel junction device, we implement the tight-binding model of Ref. \cite{Liu2013} in the Kwant transport package \cite{Groth2014} and calculate the Landauer-B\"{u}ttiker conductance. The Hamiltonian is written in the basis of $d$-orbitals and spin $\{\left(|d_z^2 \rangle, |d_z^2 \rangle, |d_z^2 \rangle \right) \otimes |\uparrow, \downarrow \rangle \}$ on a triangular lattice, and reads:

\begin{align}\label{eq_TMD}
    H =& \left( \varepsilon_0 \sum_{i,s} c_{i,s}^\dagger  c_{i,s} + \sum_{\langle i, j \rangle, s} c_{i,s}^\dagger t_{ij} c_{j,s}\right) \otimes s_0 \nonumber \\
    & + \left( \lambda L_z  \sum_{i,s} c_{i,s}^\dagger c_{i,s} \right) \otimes s_z.
\end{align}

The first term is the onsite energy with value $\varepsilon_0 = (\varepsilon_1, \varepsilon_2, \varepsilon_2)$, the second term is the nearest-neighbor hopping and the third term is the on-site spin-orbit coupling with strength $\lambda$, which in this basis only the $L_z$ component of the angular momentum operator is nonzero. Also, $s_0$ and $s_z$ are the identity and $z$-Pauli matrices, respectively, acting on the spin. The hoppings follow the symmetries of the TMD monolayer's space group, i.e. defining the hoping $t_{i1}$ as shown in the inset of Fig. \ref{fig_F1}(b), $t_{i4}$ is obtained by a reflection on the $x$ axis, and the remaining ones are obtained by three-fold rotations. The hopping $t_{i1}$ is defined as \cite{Liu2013}:
\begin{align}
    t_{i1} = \begin{pmatrix}
t_0 & t_1 & t_2\\
-t_1 & t_{11} & t_{12} \\
t_2 & -t_{12} & t_{22}
\end{pmatrix}.
\end{align}
For this work, we restrict ourselves to nearest-neighbor only, as this is enough to capture the spin-valley physics of the valence band. Noting that our results are general for any TMD showing spin-valley locking, we focus here on WSe$_2$ because it has a large energy separation between the valence band maximum of the $K, K^\prime$ points and the $\Gamma$ point. This is important as the bands at $\Gamma$ point do not show spin-valley physics and would be detrimental for the resistance switching. The values of the parameters are reported in Ref. \cite{Liu2013}, and here we just note that both the hopping amplitudes and the spin-orbit strength are of the order of few hundred meV. In Fig. \ref{fig_F1}(b) we plot the band structure of such model for WSe$_2$, where the valley-dependent spin splitting in the valence band is clearly visible.

\begin{figure}[tb]
\includegraphics[width=0.47\textwidth]{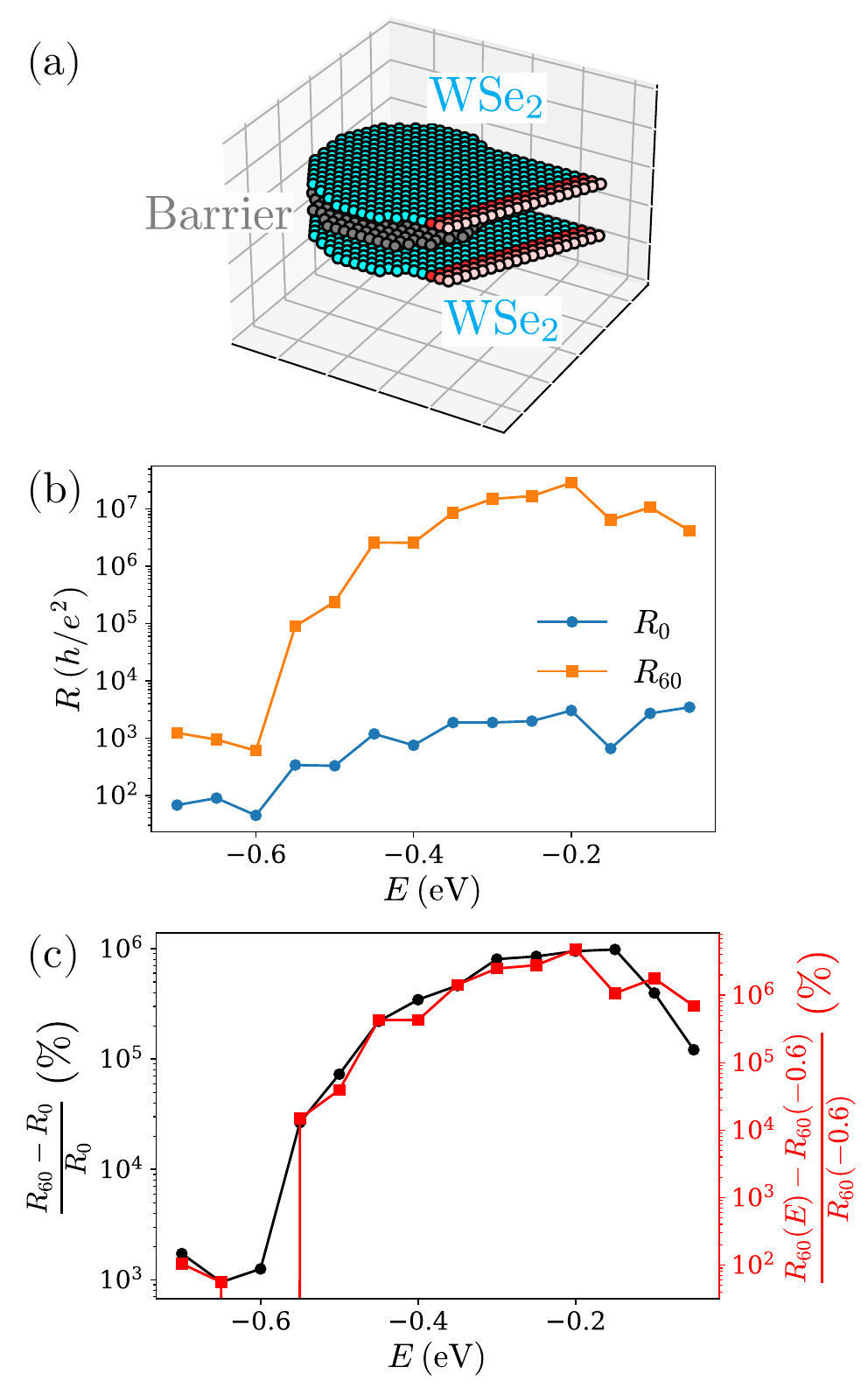}
\caption{(a) Schematics of the tunnel junction. Spheres are the tight-binding sites with colors cyan, gray and red being WSe$_2$, the barrier and three unit cells of the semi-infinite leads, respectively. (b) Resistance as a function of energy for $\theta = 0\dg$ and $\theta = 180\dg$. (c) Switching resistance \Rs as a function of energy. Black line is \Rs between $\theta = 0\dg$ and $\theta = 180\dg$, and red line is \Rs for different energies at fixed angle $\theta = 60\dg$. 
}
\label{fig_F2}
\end{figure}

Having the tight-binding for the TMDs, we now proceed in modeling the tunnel junction device, which we schematically plot in Fig. \ref{fig_F2}(a). Firstly, the bottom TMD has lattice vectors $\bm{a_1^B} = a (1,0)$, $\bm{a_2^B} = a (1/2,\sqrt{3}/2)$, with $a$ the lattice constant, whereas the top layer lattice is given by $\bm{a_1^T} = a (\cos (\theta),\sin  (\theta))$, $\bm{a_2^T} = a (\cos  (\theta)/2-\sin  (\theta)\sqrt{3}/2, \sin  (\theta)/2 + \cos  (\theta)\sqrt{3}/2)$. Then, we define a circular scattering region of diameter $L$ with common origin for both layers ($\bm{r_0^B}=\bm{r_0^T}$), meaning that the two lattices lay on top of each other at $\theta = 0$. Next, we define the tight-binding for the barrier. It is known that crystallinity and absence of scattering in the barrier helps conserve momentum and vertical transport \cite{Parkin2004, Yuasa2004, Tsymbal2007}, which is fundamental to generate $R_\text{sw}$ as explained in Fig. \ref{fig_F1}. Consequently, we start by modeling a barrier with a single-layer triangular lattice with the same orientation and origin as the bottom TMD, and stack it on top at a distance $d = a$ (this makes the TMD separation $2d$). 
The tight-binding of the barrier contains one orbital per site, with a large onsite energy of $5$ eV and small nearest-neighbor hopping of $0.02$ eV to characterize a poor conducting insulator. We choose a moderate hopping of $t_\perp = 0.05$ eV compared to the TMD hoppings to couple the barrier with WSe$_2$. Furthermore, $t_\perp$ couples equally the three TMD orbitals to the single orbital of the barrier, and a rotation operation $R(\theta)=e^{-i L_z \theta / \hbar} \otimes s_0$ acts on $t_\perp$ when the hopping is from the barrier to the top layer. Finally, for the leads, we use the same model as Eq. \eqref{eq_TMD}, and `attach' or couple one lead laterally to each WSe$_2$ layer \cite{Groth2014}. See ref. \cite{Suppmat} for additional information regarding implementation.

\textit{Results ---} We plot in Fig. \ref{fig_F2}(b) the two-terminal resistance, obtained from inverting the Landauer conductance, as a function of energy for twist angles $\theta = 0\dg$ and $\theta = 60\dg$. As expected, the resistance for $\theta = 60\dg$, $R_{60}$, is considerably larger than that of aligned TMDs, $R_0$. Both resistances decrease similarly with moving away from the valence band maximum because there are more states available for transport. However, around $E \sim -0.45$ eV, $R_{60}$ starts dropping at a faster rate and becomes much more similar to $R_{0}$ after $E \sim -0.55$ eV. This occurs because the band with opposite spin at each valley becomes populated at $E \sim -0.45$ eV while states at the $\Gamma$ point appear at $E \sim -0.55$ eV. From these resistances, we plot \Rs in Fig. \ref{fig_F2}(c), revealing a strikingly switching resistance larger than $10^6 \%$. This value compares well with recent experimental tunnel magnetoresistance in two-dimensional ferromagnetets \cite{Song2018, Klein2018, Wang2018NC} with the advantage that TMDs are not limited to operate below the Curie temperature \cite{Kurebayashi2022}. 
As mentioned above, $R_{60}$ changes abruptly near $E \sim -0.45$ eV. Therefore, at a fixed angle of $\theta = 60\dg$, one can define another switching resistance as a function of energy where $R_\text{low}$ is taken at some energy below $E \sim -0.45$ eV, e.g. $(R_{60}(E) - R_{60}(-0.6)) / R_{60}(-0.6) \%$. We plot this quantity as well in Fig. \ref{fig_F2}(c), and notice that \Rs also surpasses $10^6 \%$, implying that electrical tuning of the Fermi level could also be used to obtain large switching resistances.

\begin{figure}[tb]
\includegraphics[width=0.47\textwidth]{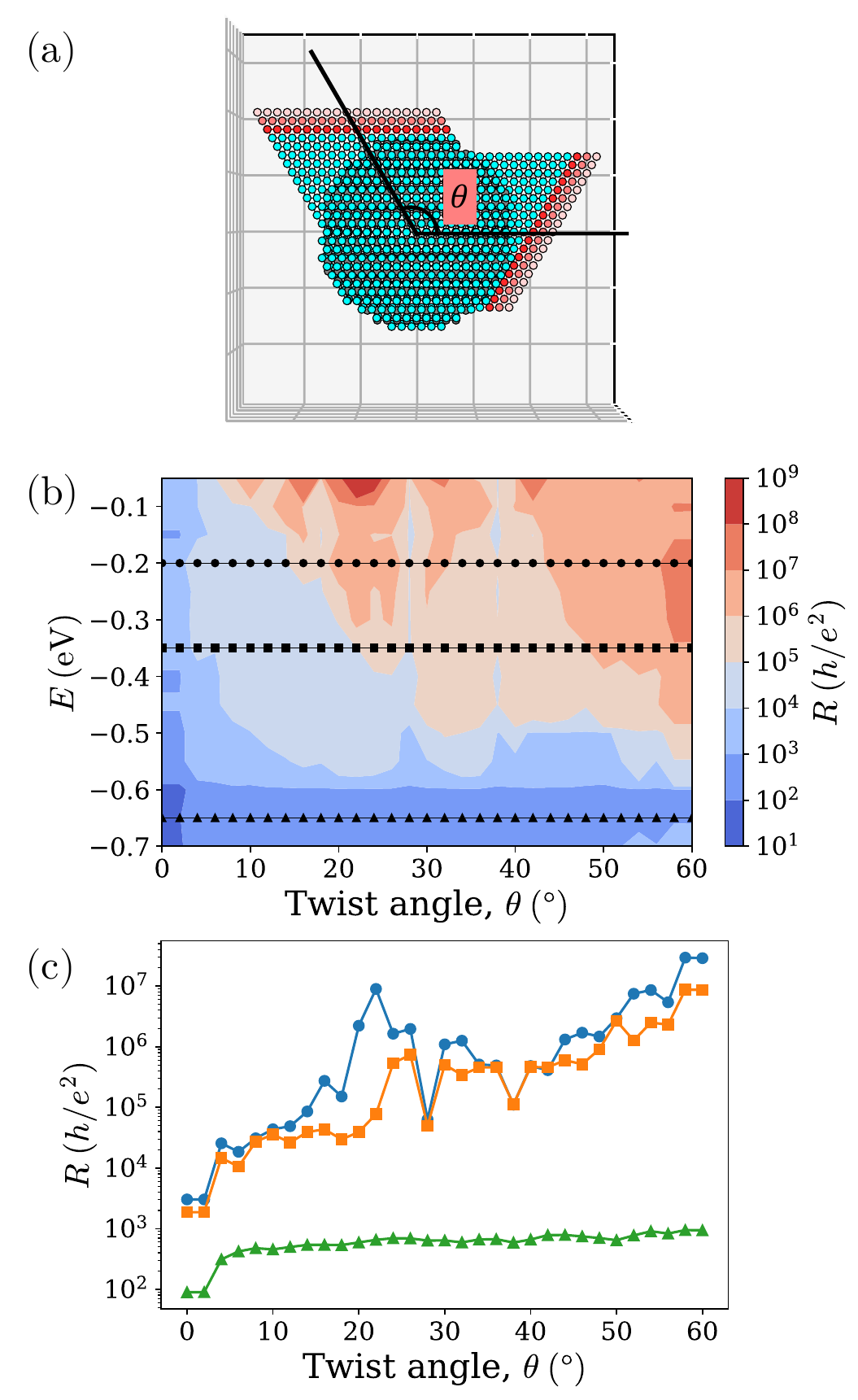}
\caption{(a) Schematics of the tunnel junction with arbitrary twist angle $\theta$. (b) Resistance as a function of energy and twist angle for the same device from Fig. \ref{fig_F2}. (c) Resistance as a function of twist angle for three selected energies marked with lines in panel (b).}
\label{fig_F3}
\end{figure}

Although the basic concept of this work compares the resistance between TMD bilayers with the same and opposite sign of spin-valley locking, one may wonder what happens for twist angles between 0 and 60 degrees. For example, for $\theta = 30\dg$ the Fermi surface pockets of the two layers may not coincide at all in the Brillouin zone, thus making the device even more insulating. To elucidate such nontrivial trend, we calculate the resistance as a function of both energy and twist angle, and plot the results in Fig. \ref{fig_F3}. The general trend is that the resistance increases with increasing twist angle, but there are additional features. For instance, the resistance does not increase monotonically from $0\dg$ but instead it remains quite constant for small angles and then suddenly increases around $\theta \sim 4\dg$. After that, the growth is reasonable constant with twist angle, with some differences depending on the energy \cite{Suppmat}. Logically, for larger energies where both spin species are present at each valley, the resistance barely changes with the angle.

\begin{figure}[tb]
\includegraphics[width=0.47\textwidth]{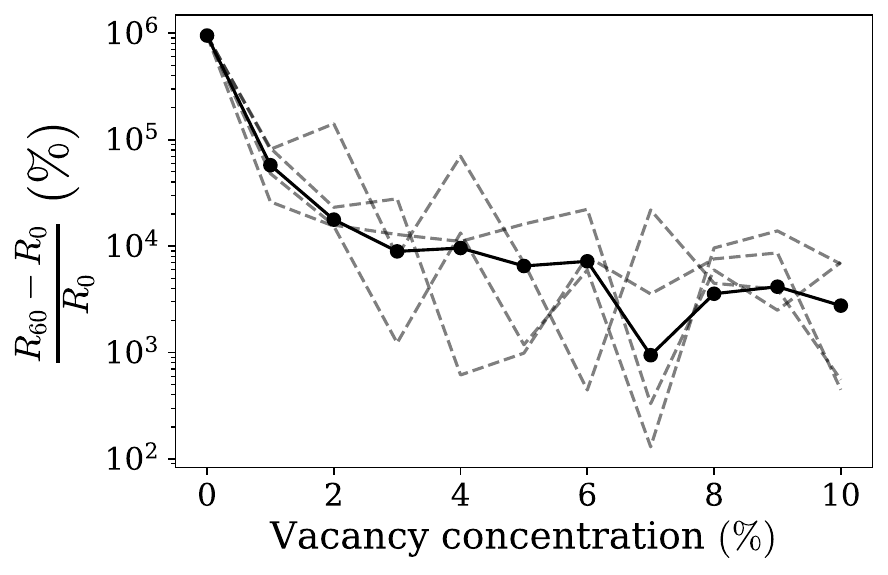}
\caption{(a) Switching resistance \Rs as a function of vacancy concentration at $E = -0.2$ eV. Solid line is the average performed over 5 disorder realizations (shown in dashed, paler color).}
\label{fig_F4}
\end{figure}

Previously, we mentioned that scattering in the metal-barrier interface is important as that can couple states with different momentum (and same spin) and reduce \Rs. Importantly for us, not all kinds of disorder will affect \Rs in the same manner. Long-range disorder will induce scattering involving small momentum changes, i.e. intra-valley scattering. This type of disorder should not severely impact \Rs. On the other hand, short-range scatterers, such as adatoms or vacancies, can mix states far away in the Brillouin zone and therefore produce inter-valley scattering that allow vertical transport even when the TMDs are not aligned. To unveil the dependence of inter-valley scattering to the switching resistance, we model vacancies in our tight-binding model \cite{Suppmat,Uppstu2014, Fan2014}. 
In Fig. \ref{fig_F4}, we plot \Rs as a function of vacancy concentration at $E = -0.2$ eV. Clearly, \Rs quickly drecreases with increasing concentration. However, experiments have shown that typical defect concentration in TMDs ranges $\sim 3 \%$ \cite{Hong2015, Roy2018, Gali2020}, and for that value, \Rs is still large with value $\sim 10^4 \%$. This suggests that ultraclean TMD monolayers are not needed to experience the \Rs presented here.


\textit{Discussion ---} We have revealed a spin-dependent switching resistance in a system without magnetic order. Such hitherto unprecedented phenomena originates from the effective tunning of SOC in TMD bilayers with twist angle.  Carefully tuning the twist angle between TMD monolayers allows to control the relative sign of the spin-valley locking between layers, achieving in this way low and high resistance states in vertical transport depending on the relative orientation of the TMDs. The efficiency of the resistance switch is above $10^6 \%$, making it comparable to state-of-the-art two-dimensional magnetic tunnel junctions.

Motivated by the fascinating physics of moir{\'e} materials \cite{Andrei2021, Cao2018}, nanoscale control of twist angles in van der Waals heterostructures has seen a major development in recent years \cite{RibeiroPalau2018, Inbar2023}. Specifically, the multiple works demonstrating sliding ferroelectricity in TMDs \cite{Wang2022TMD, Weston2022, Rogee2022} suggest that our proposal device should be within current experimental reach, while the quantum twisting microscope offers the perfect setup to measure tunnel currents as a function of twist angle in TMD tunnel junctions \cite{Inbar2023}. Because of the large spin-orbit and spin splitting, the resistance switching should persist at room temperature, which should facilitate its experimental identification. Overall, our work puts forward a pioneering way to create next-generation memory and electronic devices based on twist angle engineering.




\begin{acknowledgments}
M. V. is grateful to Daniel E. Parker, Tiancong Zhu, Michael F. Crommie and Andrew D. Kent for stimulating discussions. M.V. was supported as part of the Center for Novel Pathways to Quantum Coherence in Materials, an Energy Frontier Research Center funded by the US Department of Energy, Office of Science, Basic Energy Sciences.
\end{acknowledgments}

%

\end{document}